\documentclass[%
reprint,
showpacs,preprintnumbers,
superscriptaddress,
]{revtex4-1}

\usepackage{graphicx}
\usepackage{dcolumn}
\usepackage{bm}
\usepackage{natbib}
\usepackage{hyperref}
\hypersetup{
    colorlinks=true,
    linkcolor=blue,
    filecolor=magenta,      
    urlcolor=blue}
\usepackage[mathlines]{lineno}
 
\begin{document}

\preprint{APS/123-QED}

\title{Influence of vibrational excitation on surface diffuseness 
of inter-nuclear potential: study through heavy-ion quasi-elastic 
scattering at deep subbarrier energies}



\author{Gurpreet Kaur}
\email{gkaur.phy@gmail.com}
\author{B.R. Behera}
\affiliation{Department of Physics, Panjab University, Chandigarh-160014, India}

\author{A. Jhingan}
\author{P. Sugathan}
\affiliation{Inter University Accelerator Centre, Aruna Asaf Ali Marg, New Delhi-110067, India}

\author{K. Hagino}
\affiliation{Department of Physics, Tohoku University, Sendai 980-8578, Japan}
\affiliation{
Research Center for Electron Photon Science, Tohoku University, 1-2-1 Mikamine, Sendai 982-0826, Japan}
\affiliation{
National Astronomical Observatory of Japan, 2-21-1 Osawa,
Mitaka, Tokyo 181-8588, Japan}

\begin{abstract}
We discuss the role of channel coupling in the surface 
properties of an inter-nuclear potential for heavy-ion 
reactions. 
To this end, we analyze the experimental quasi-elastic cross sections 
for the $^{12}$C + $^{105,106}$Pd and $^{13}$C + $^{105,106}$Pd systems 
using the coupled-channels approach by including the vibrational 
excitations in the target nuclei. 
While 
earlier studies have reported a negligible influence of vibrational 
excitation on the surface diffuseness parameter for spherical systems, 
we find a significant effect for the C+Pd systems. 
Our systematic study also reveals influence of transfer couplings 
on the surface diffuseness parameter. 
\end{abstract}

          
\date{\today}

\maketitle

\section{\label{sec:level1}{INTRODUCTION}}

Even though fusion cross sections reflect the dynamical behavior of the nuclei involved in the fusion \cite{Steadman}, apparently it is incapable to get insights into the interaction mechanism between nuclei simply by measuring fusion cross sections. That is, theoretical calculations, such as coupled-channels calculations (either semi-classical or quantal), are indispensable in order to understand the fusion dynamics \cite{Dasgupta1,BT98,HT12,Back14}. These calculations, in comparison with the experimental data, address the influence of coupling between the relative motion and the nuclear intrinsic degrees of freedom, and thus 
the interplay between nuclear structure and reaction dynamics \cite{Stefanini,Dasgupta2}. As a result, the coupled-channels formalism has become a powerful and standard tool in order to interpret experimental fusion cross sections\cite{HT12}. 

One of the most important ingredients of such calculations is an inter-nuclear potential. Various types of potential, such as a double folding potential \cite{Hagino1,Gonttchar} and a phenomenological Woods-Saxon potential \cite{wood}, have been employed. Among these, the Woods-Saxon form, described by the depth parameter, $V_0$, the radius parameter, $r_0$, and the surface diffuseness parameter, $a_0$, has gained popularity due to its simplicity and ability to reproduce many experimental results. While representing the Woods-Saxon potential, its depth and radius parameters can be mutually adjusted to reproduce the Coulomb barrier of the system once the surface diffuseness parameter is fixed. Obtaining the quantitative measure for the surface diffuseness parameter is therefore of crucial
 importance, as the curvature of the Coulomb barrier is largely determined by the surface diffuseness parameter, which by definition reflects the surface property of the internuclear potential. 

Three different methods have been used in literature in order to extract a diffuseness parameter from experimentally measured data. The first method is to analyze elastic scattering cross sections either with an optical potential model or with a coupled-channels approach. This methods has established a value of surface diffuseness to be around 0.6 fm for most of the systems \cite{elastic}. The second method is to use high precision fusion cross section. It has turned out that this method leads to a much larger value of surface diffuseness parameter, ranging from 0.75 to 1.5 fm, than the first method \cite{newton1,newton2}. Recently, Hagino et al. have proposed the third method, which uses a quasi-elastic (QE) excitation function at large backward angles \cite{Hagino2}. They have suggested that at energies well below the Coulomb barrier, a QE excitation function is sensitive mainly to the surface property of nuclear potential and can be used to obtain quantitatively the surface diffuseness parameter. This method leads to the diffuseness parameter for spherical systems to be around 0.6 fm \cite{Washiyama}, while somewhat larger values of the diffuseness parameter have been obtained for systems involving deformed nuclei.

The different values of surface diffuseness parameter, obtained for spherical and deformed systems, lead to a realization that not only obtaining the quantitative measure of diffuseness but also understanding how the coupling affects the surface diffuseness is of importance. For this purpose, several studies have been carried out in aiming at understanding the effects of different couplings on the surface diffuseness parameter. Gasques et al. \cite{Gasques} have performed both the single-channel (sc) and the coupled-channels (cc) calculations for $^{32}$S+$^{208}$Pb, $^{197}$Au, $^{186}$W, and $^{170}$Eu systems and concluded that the rotation coupling shows a significant influence on the diffuseness parameter value. See also Ref. \cite{Evers08}. Washiyama et al. \cite{Washiyama} have analyzed the QE measurements for $^{16}$O, $^{32,34}$S+ $^{208}$Pb systems with the single-channel calculations. They did not obtain any enhancement in the value of surface diffuseness parameter, and thus a significant effect of vibrational coupling on the diffuseness parameter was not observed. Studies on many other systems, such as $^{16,17,18}$O+ $^{92}$Mo \cite{92Mo}, have also shown similar results. Hence, it has generally been considered that the vibrational excitation has a marginal effect on the surface diffuseness parameter. 

The aim of this paper is to investigate further the effect of vibrational excitations on the surface diffuseness parameter. As in the rotational excitations, one may expect that the effect becomes more significant when the vibrational excitation energy is small. Furthermore, we also investigate the effect of transfer coupling, which has been established to play an important role in the fusion dynamics \cite{Jiang14}. To the best of our knowledge, no study has reported concerning how the transfer coupling affects the diffuseness parameter. For these purposes, we re-analyze the experimental data for $^{12,13}$C+$^{105,106}$Pd systems reported in Ref. \cite{QEdata} using the coupled-channels approach. 

The paper is organized as follows. In Sec. II, we first discuss the characteristic feature of the $^{12,13}$C+$^{105,106}$Pd systems with respect to the vibrational and the transfer couplings. We then detail the procedure to extract the surface diffuseness parameter through the analyses of the quasi-elastic data. In Sec. III, we present the results of our analyses and discuss the role of vibrational and transfer couplings. Finally, we summarize the paper in Sec. IV.

\section{Method}

\subsection{Systems}

In order to study the effect of vibrational coupling on a surface diffuseness parameter, we re-analyze the experimental quasi-elastic scattering for the $^{12,13}$C+$^{106,106}$Pd systems, for which the data have been available in Ref. \cite{QEdata}. While Ref. \cite{QEdata} focused on extracting the barrier distributions, 
our aim in this paper is to extract the surface diffuseness parameter using the deep-subbarrier data. We choose these systems as the target nuclei, $^{105,106}$Pd, exhibit low-energy vibrational excitations ($E^*$ $\approx$ 0.5 MeV for the first excited state in $^{104,106}$Pd). The projectile, $^{13}$C, along with the chosen target, is an ideal candidate to study the effect of transfer coupling due to the following characteristics:
\begin{enumerate}
\item Positive Q-values for one-neutron transfer channels (see Table \ref{tab:Q-value}), which are necessary for a system to exhibit significant influences of a neutron transfer coupling \cite{Jiang14,Qvalue}.
\item The existence of the weakly bound valence nucleon, which ensures the transfer of neutron. 
\item Vibrational couplings in the target nuclei, which are in general weaker as compared to rotational couplings so that the effect of transfer may not be significantly masked. 
\end{enumerate}

In our systematic study, we carry out the coupled-channels calculations by including the vibrational couplings in the target nuclei in order to understand the influence of the vibrational coupling on the surface diffuseness parameter. Moreover, since the $^{12}$C + $^{105,106}$Pd systems have a negative Q-value for the neutron pick-up reactions (see Table \ref{tab:Q-value}), a comparison with the $^{13}$C + $^{105,106}$Pd systems will elucidate the role of transfer coupling in the surface diffuseness parameter.

\begin{table}[bt]
\begin{center}
\caption{\label{tab:Q-value}
The Q-value for the neutron transfer channels for the $^{12,13}$C+$^{105,106}$Pd systems, given in units of MeV. 
Here the negative and positive signs correspond to the pick-up and stripping reactions, respectively.}
\begin{ruledtabular}
\begin{tabular}{ccccc}
System &  $(-1n)$ & $(-2n)$ & $(+1n)$ & $(+2n)$ \\
\hline
\vspace{+0.075cm}
$^{12}$C + $^{105}$Pd & $-$9.161 & $-$15.744 & $-$2.148 & $-$3.953\\
\vspace{+0.075cm}
$^{12}$C + $^{106}$Pd & $-$12.185 & $-$16.082 & $-$4.615 & $-$3.532\\
\vspace{+0.075cm}
$^{13}$C + $^{105}$Pd & $+$4.615 & $-$7.571 & $+$1.082 & $-$7.681\\
\vspace{+0.075cm}
$^{13}$C + $^{106}$Pd & $+$1.59 & $-$7.908 & $-$1.385 & $-$7.261\\
\end{tabular}
\end{ruledtabular}
\end{center}
\end{table}

\subsection{Procedure}

To perform a systematic study, the single-channel and coupled-channel calculations have been performed using a scattering version of the {\tt CCFULL} program \cite{ccfull}. For the coupled-channels calculations, we have included the double quadrupole phonon excitations in the target nuclei in the harmonic oscillator limit. 
The deformation parameter and the excitation energy for $^{106}$Pd are given by $\beta$=0.229 and $E^*$=0.512 MeV, respectively \cite{QEdata}. For the $^{105}$Pd nucleus, we have followed Ref. \cite{QEdata} and have taken the average in the adjacent nuclei, that is, $^{104}$Pd and $^{106}$Pd, which leads to $\beta$=0.219 and $E^*$=0.534 MeV. 

The nuclear potential used in the calculations has a real and an imaginary components, both of which are assumed to have a Woods-Saxon form. The imaginary part simulates a compound nucleus formation. We have chosen the strength to be large enough so that the flux does not reflect inside the barrier once the barrier is overcome. In the calculations, we have used an imaginary potential with the depth parameter of 30 MeV, the radius parameter of 1.0 fm, and the diffuseness parameter of 0.3 fm. This choice of parameters confines the imaginary potential well inside the Coulomb barrier with a negligible strength in the surface region. 
As long as the imaginary potential is confined inside the Coulomb barrier with a large strength, the results are insensitive to the parameters of the imaginary part. For the real part of the nuclear potential, the potential depth $V_0$ is fixed to be 185 MeV. The value of radius parameter $r_0$ is then adjusted for a particular value of the diffuseness parameter such that the Coulomb barrier height $V_B$ for each system becomes the same as that for the Bass potential \cite{bass}. This is possible because the effect of variation in $V_0$ and $r_0$ on the Coulomb barrier height compensates with each other in the surface region. That is, for a given value of diffuseness parameter, the results do not significantly depend upon the actual choice of $ V_0$, as long as the same barrier height $V_B$  is maintained.

To ensure that the barrier height for the single-channel and the coupled-channels calculations corresponds to the same value, we have slightly readjusted the potential parameters for the coupled-channels calculations by using the fusion cross sections at energies above the barrier. The following steps have been taken for this purpose: 

\begin{enumerate}
\item For a chosen and fixed value of the diffuseness parameter $a_0$, and with the depth $V_0$ = 185 MeV, the value of the radius parameter $r_0$ is determined such that the Coulomb barrier energy $V_B$ reproduces the Bass barrier. 

\item The fusion cross sections, $\sigma_{\rm fus}^{(sc)}$, are calculated using the single-channel calculation. 

\item The full coupled-channels calculations are then performed to obtain the fusion cross sections in the presence of the channel couplings, $\sigma_{\rm fus}^{(cc)}$.  In general, even at energies above the barrier, these fusion cross sections are different from $\sigma_{\rm fus}^{(sc)}$ due to the potential renormalization \cite{Hagino3}. By slightly adjusting the radius parameter, $r_0$, we match the fusion cross sections $\sigma_{\rm fus}^{(cc)}$ 
at energies well above the barrier with $\sigma_{\rm fus}^{(sc)}$. This results in a set of potential parameters ($V_0$, $r_0$, $a_0$) that reproduces the fusion 
barrier height by taking into account the couplings to the intrinsic states with the coupled-channels calculations. 

\item For every set of nuclear potential parameters, both for the single-channel and the coupled-channels analyses, the quasi-elastic scattering cross sections are computed. For the single-channel calculation, the quasi-elastic cross section corresponds simply to the elastic scattering cross section. On the other hand, for the coupled-channels calculations, the quasi-elastic cross sections correspond to a sum of elastic and inelastic cross sections.

\end{enumerate}

In order to find the best fitted value of the diffuseness parameter, 
the chi square $\chi^{2}$ method has been utilized. To this end, the data with $d\sigma_{\rm qel}/d\sigma_{\rm R}>1$, where $d\sigma_{\rm qel}/d\Omega$ and $d\sigma_{\rm R}/d\Omega$ are quasi-elastic and the Rutherford cross sections, respectively, 
have been excluded from the fitting procedures, 
even though they are included in the figures for completeness. 
The uncertainty in the optimum value of $a_0$ has been calculated using the following procedure. 
For the $\chi^{2}_{\rm min}$ value corresponding to the best fit value of the diffuseness parameter, 
the quantity $(\chi^{2}_{\rm min}$ + $\chi^{2}_{\rm min}$ $/n)$ was calculated, where $n$ denotes 
the number of degrees of freedom. 
The intersection of this quantity with the $\chi^{2}$ envelope gives 
the two values $a_0^-$ and $a_0^+$ defining the error in $a_0$.

For the $^{12}$C + $^{105,106}$Pd and $^{13}$C + $^{105,106}$Pd systems, the QE excitation functions have 
been measured at 165$^\circ$ in the laboratory frame as reported in Ref. \cite{QEdata}. 
To be consistent, 
all the calculations have been carried out at same scattering angle. 
In order to ensure that the calculations are properly scaled according to the available data, 
the calculated ratio of the quasi-elastic to the Rutherford cross sections are analyzed and 
plotted as functions of the effective energy defined as \cite{eff1,eff2}, 
\begin{equation}
E_{\rm eff} ={\frac{2E_{\rm c.m.}}{1 + {\rm cosec}\left(\frac{\theta_{\rm c.m.}}{2}\right)}}
\end{equation} 
where $E_{\rm c.m.}$ and $\theta_{\rm c.m.}$ are energy and scattering angle in the 
center of mass frame, respectively. 
This corrects for the ``angle dependent" centrifugal effects by making 
$\sigma_{\rm qe}({E_{\rm eff}, 180^\circ) \approx \sigma_{\rm qe} 
(E_{\rm c.m.}, \theta_{\rm c.m.}})$.

\section{Results and Discussion}

\subsection{Single-channel calculations}

\begin{figure*}[pt]
\centering
\includegraphics[width=.46\linewidth]{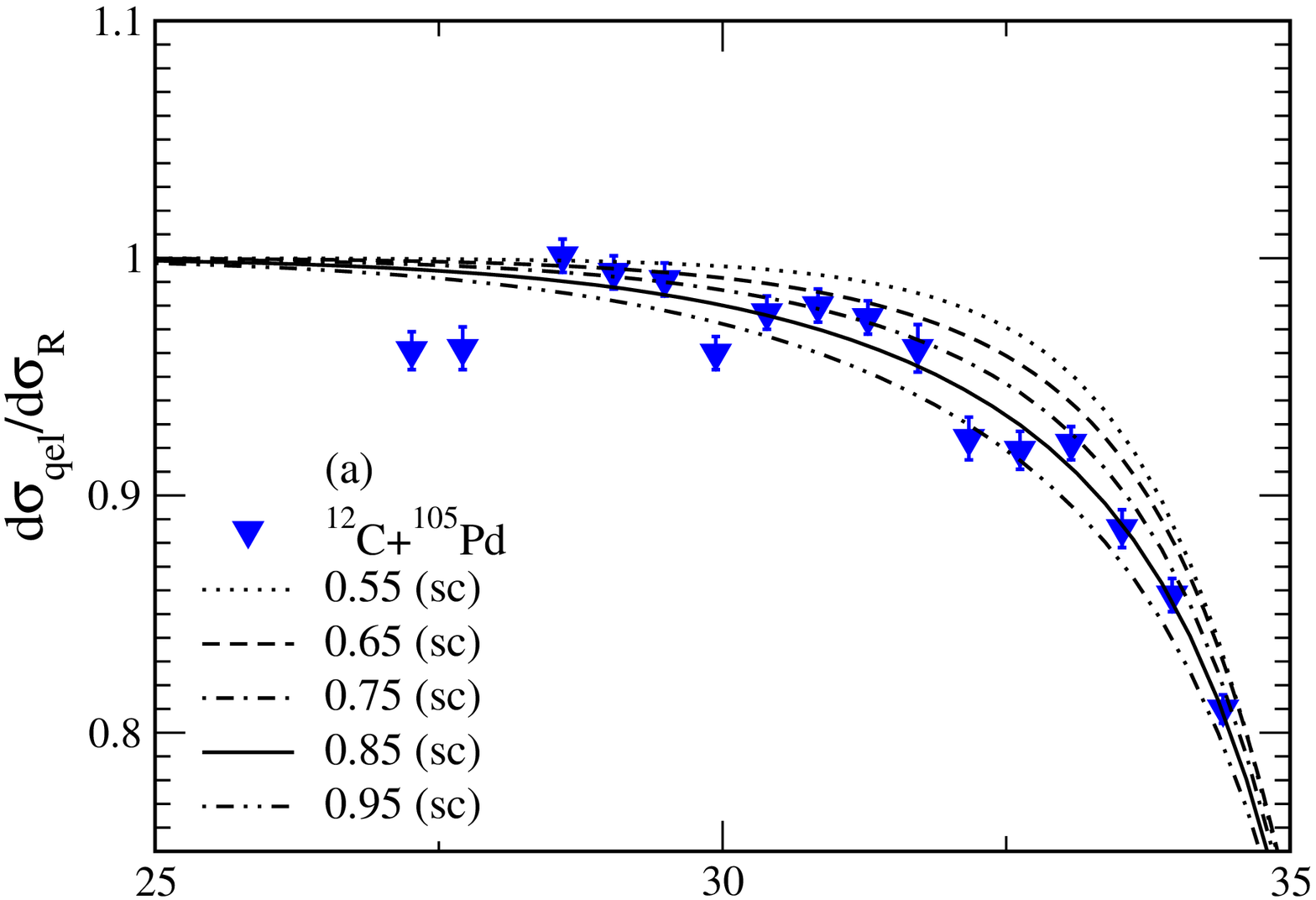}
\includegraphics[width=.43\linewidth]{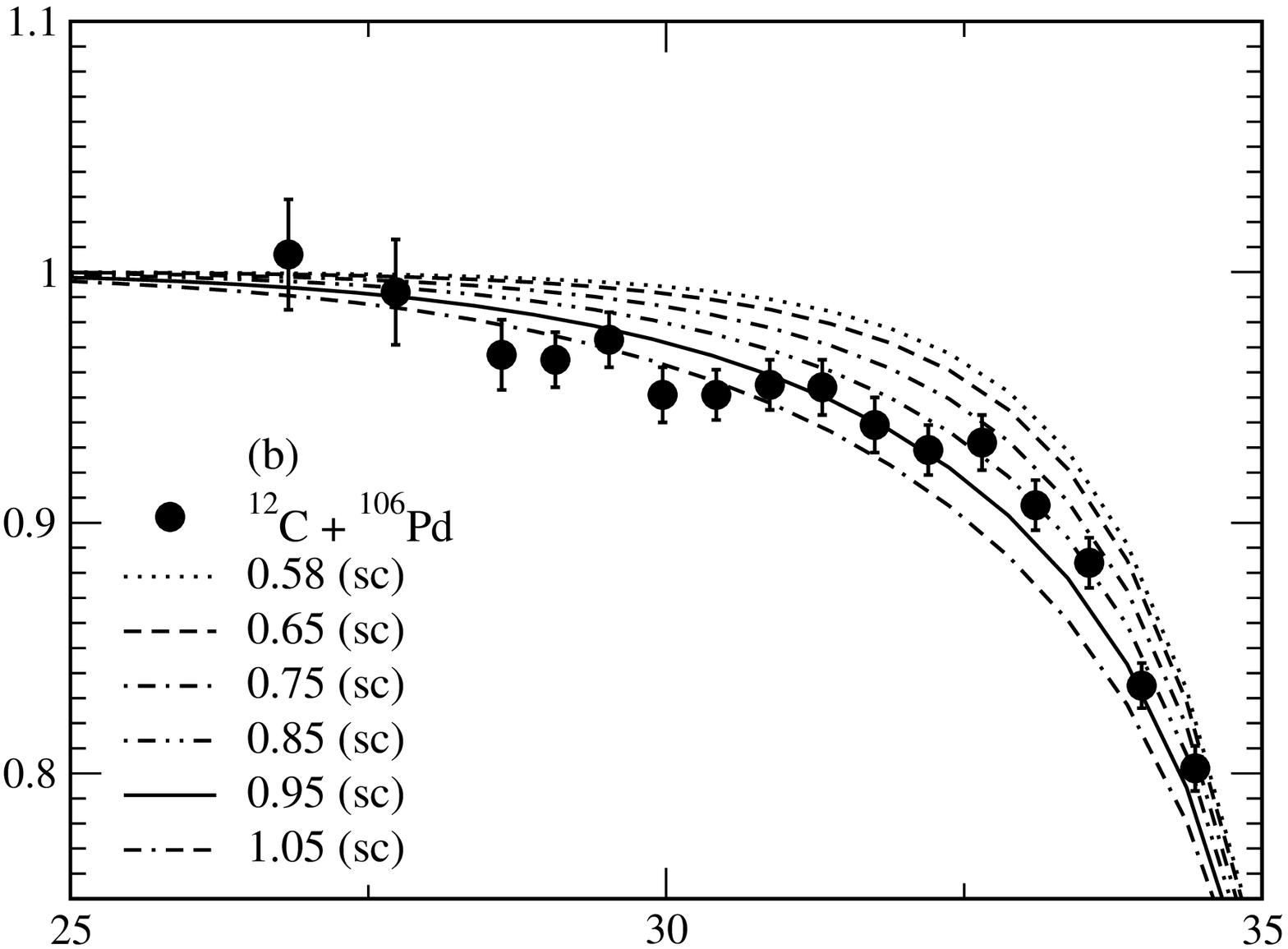}
\includegraphics[width=.46\linewidth]{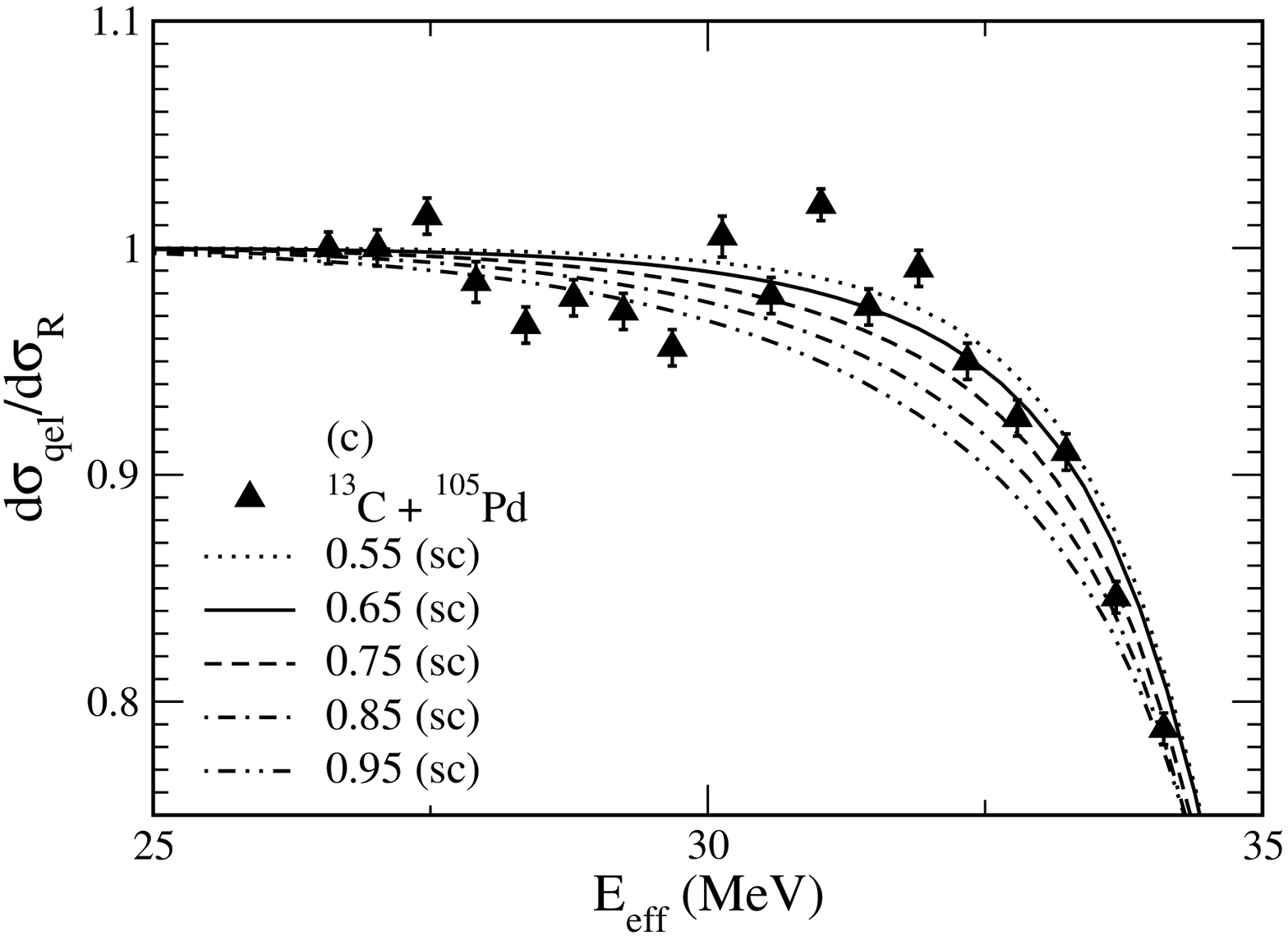}
\includegraphics[width=.43\linewidth]{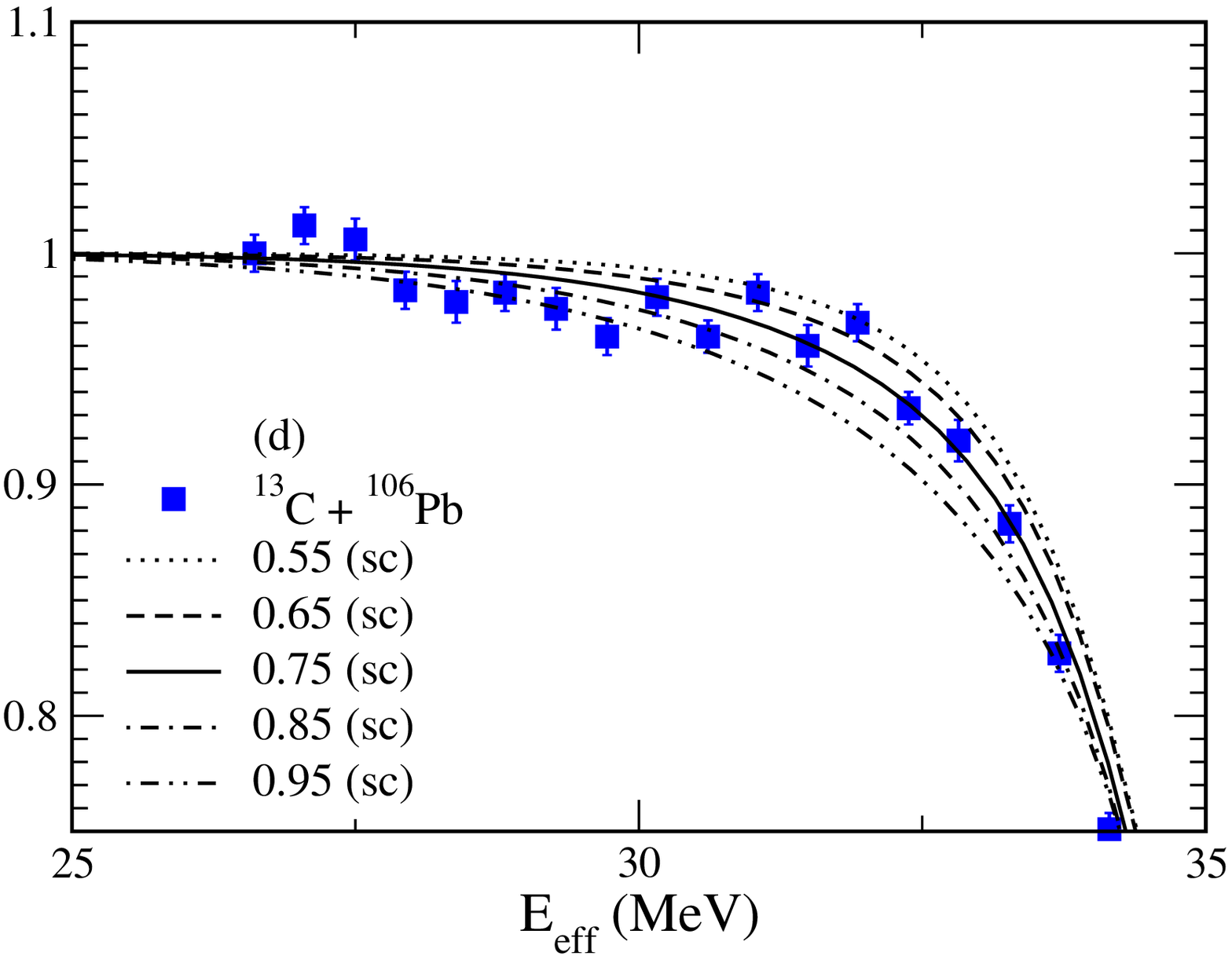}
\caption{\label{fig:sc}
(Color online) Comparisons of the single-channel calculations for the 
quasi-elastic excitation function obtained with several values of the surface diffuseness parameter, $a_0$, in the nuclear potential. 
The panels (a), (b), (c), and (d) are for the 
$^{12}$C + $^{105}$Pd, $^{12}$C + $^{106}$Pd, 
$^{13}$C + $^{105}$Pd and $^{13}$C + $^{106}$Pd systems, respectively. 
The experimental data are taken from Ref. \cite{QEdata}. }
\end{figure*}

\begin{figure}[pt]
\includegraphics[width=.68\linewidth]{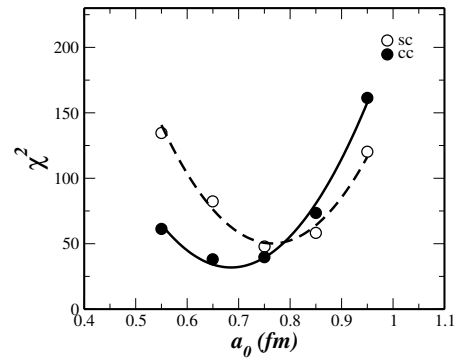}
\caption{\label{fig:chi2}
The chi square ($\chi^2$) for the $^{13}$C + $^{106}$Pd system as a function of the surface diffuseness 
parameter, $a_0$. 
The open and filled symbols represent the results of 
the single-channel and coupled-channel calculations, respectively.}
\end{figure}

Let us now numerically extract the surface diffuseness parameter from the experimental data for the quasi-elastic 
scattering. We have first performed the single-channel calculations without including the inelastic excitations 
of the target nuclei. 
The value of $a_0$ has been extracted using the procedure explained in the previous section. 
After the chi-square fitting, 
the best fitted value of $a_0$ for the $^{12}$C + $^{105}$Pd and $^{12}$C + $^{106}$Pd systems 
have been found to be 0.80$\pm$0.04 fm and 0.94$\pm$0.07 fm, respectively. 
The quasi-elastic cross sections obtained with several values of the surface diffuseness parameter are 
shown in Figs. \ref{fig:sc} (a) and \ref{fig:sc} (b) for the  
$^{12}$C + $^{105}$Pd and $^{12}$C + $^{106}$Pd systems, respectively. 
The chi-square fit for the $^{12}$C + $^{105}$Pd system is shown in Fig. \ref{fig:chi2}. 
Similarly, for the $^{13}$C + $^{105}$Pd (shown in Fig. \ref{fig:sc} (c)) and 
$^{13}$C + $^{106}$Pd systems (shown in Fig. \ref{fig:sc} (d)), 
the best fitted values of $a_0$ after minimizing the $\chi^2$ have been found to be 
0.64$\pm$0.05 fm and 0.76$\pm$0.04 fm, respectively.

The optimum values of the surface diffuseness parameter are summarized in Table \ref{tab:a}. 
We notice that these values are significantly larger than the ``standard value'' of around 0.6 fm (obtained from elastic scattering cross sections), which are 
in a similar situation as in systems with a deformed target. 

\begin{table}[bt]
\begin{center}
\caption{\label{tab:a}
The optimum value of the surface diffuseness parameter, $a_0$, obtained with the single-channel 
and the coupled-channels calculations. Those values 
are given in units of fm.} 
\begin{ruledtabular}
\begin{tabular}{ccc}
System &  single-channel  & coupled-channels \\
\hline
\vspace{+0.075cm}
$^{12}$C + $^{105}$Pd & 0.80$\pm$0.04 &  0.69$\pm$0.04 \\
\vspace{+0.075cm}
$^{12}$C + $^{106}$Pd & 0.94$\pm$0.07 & 0.78$\pm$0.05 \\
\vspace{+0.075cm}
$^{13}$C + $^{105}$Pd & 0.64$\pm$0.05 & 0.60$\pm$0.05 \\
\vspace{+0.075cm}
$^{13}$C + $^{106}$Pd & 0.76$\pm$0.04 & 0.68$\pm$0.03 \\
\end{tabular}
\end{ruledtabular}
\end{center}
\end{table}

\subsection{Coupled-channels calculations}

\begin{figure*}[pt]
\centering
\includegraphics[width=.46\linewidth]{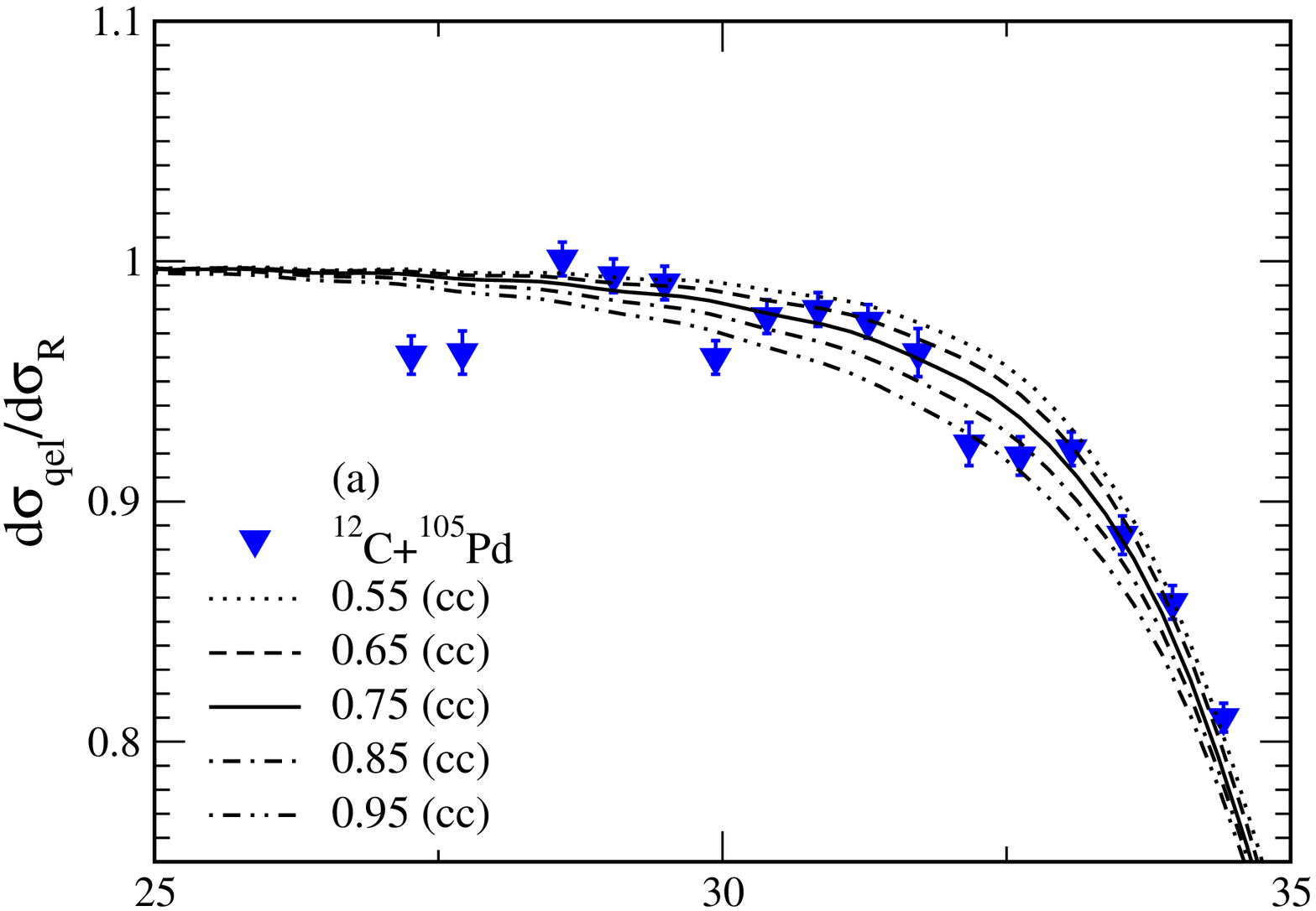}
\includegraphics[width=.43\linewidth]{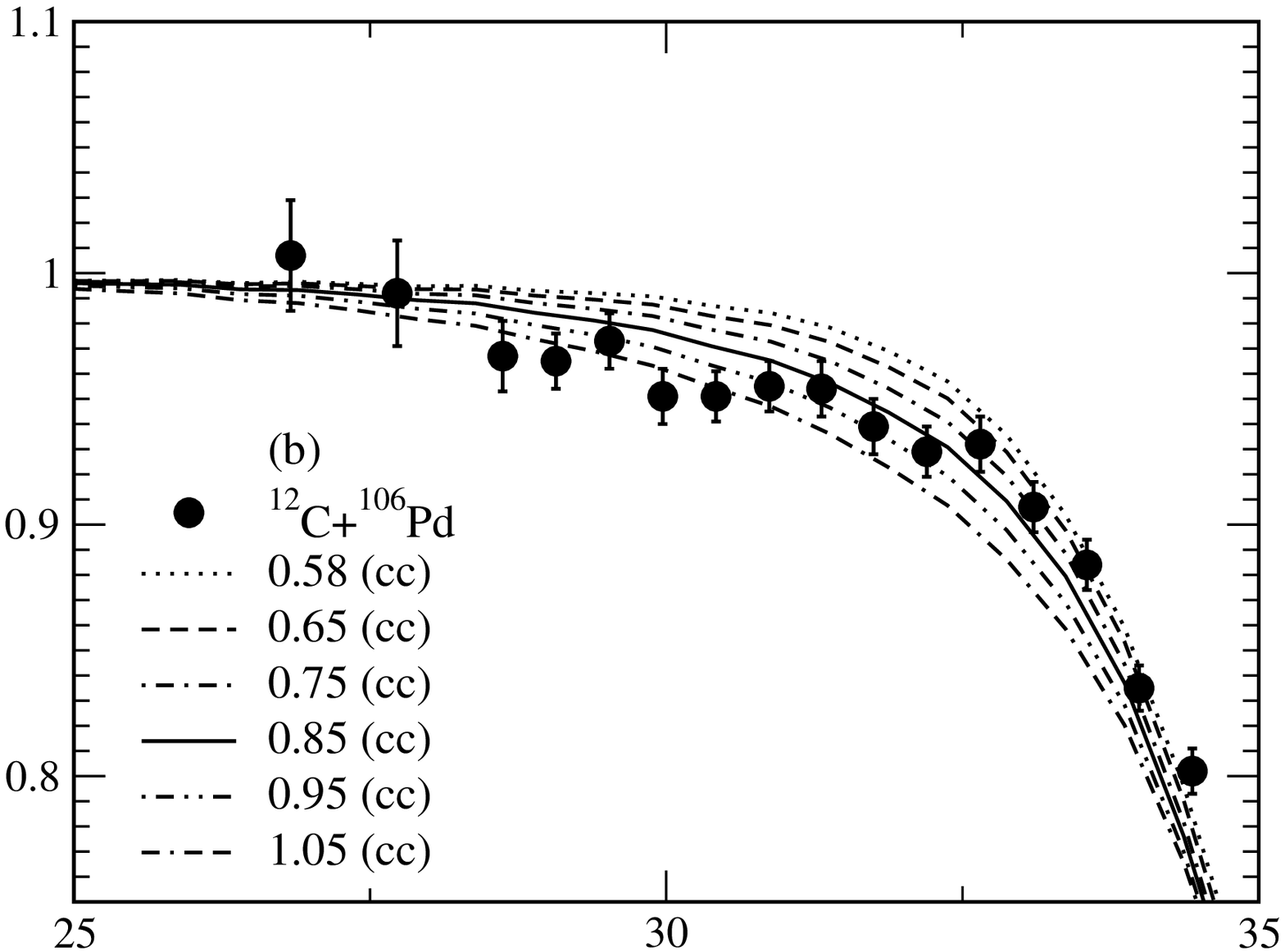}
\includegraphics[width=.46\linewidth]{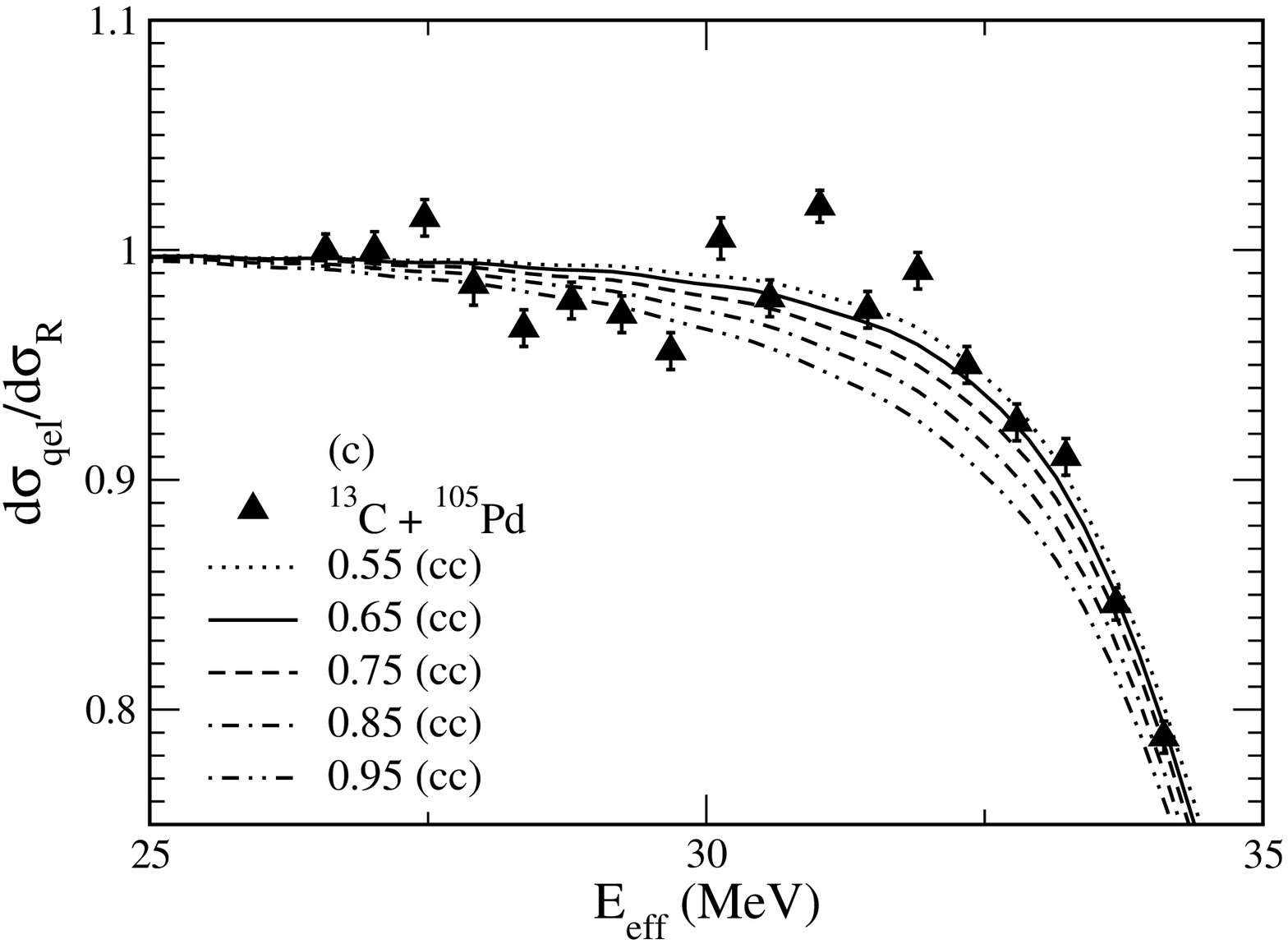}
\includegraphics[width=.43\linewidth]{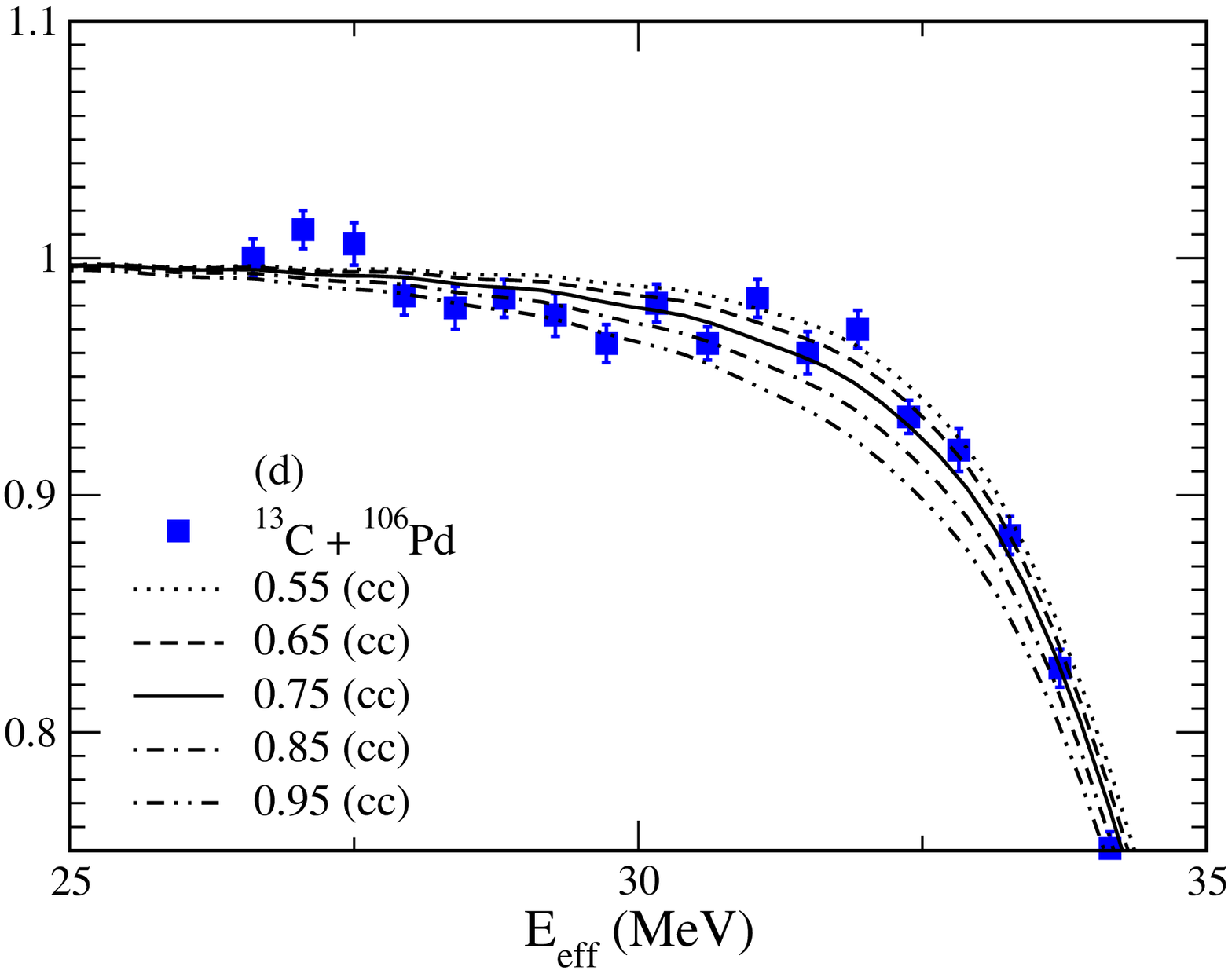}
\caption{\label{fig:cc}(Color online) 
Same as Fig. \ref{fig:sc}, but with the coupled-channels calculations.}
\end{figure*}

In order to investigate whether the large values of surface diffuseness parameter obtained with 
the single-channel calculations are due to the neglect of channel coupling effects, 
we have next performed the coupled-channels calculations 
including the vibrational excitations in the target nuclei, $^{105,106}$Pd. 
The value of $a_0$ has been varied and the best fitted value has been obtained after the 
$\chi^2$ minimization. 
The comparison of the coupled-channels calculations with several values of $a_0$ with the experimental 
data is shown in Fig. \ref{fig:cc}. 
The best fitted value of $a_0$ for the $^{12}$C + $^{105}$Pd (shown in Fig. \ref{fig:cc} (a)) and 
the $^{12}$C + $^{106}$Pd systems (shown in Fig. \ref{fig:cc} (b)) have been found to be 0.69$\pm$0.04 fm 
and 0.78$\pm$0.05 fm, respectively. 
Similarly, the best fitted value of $a_0$ for the $^{13}$C + $^{105}$Pd (shown in Fig. \ref{fig:cc} (c)) 
and the $^{13}$C + $^{106}$Pd systems (shown in Fig. \ref{fig:cc} (d)) 
are 0.60$\pm$0.05 fm and 0.68$\pm$0.03 fm, respectively.
Those values are summarized in Table \ref{tab:a}. 

It is apparent from Table \ref{tab:a} that the diffuseness parameter decreases in the coupled-channels calculations as compared to the single-channel calculations with inert target nuclei. This observation is similar to that observed in the case of rotational coupling \cite{Gasques}. Notice that earlier studies with systems such as $^{32}$S + $^{208}$Pb did not show any influence of a vibrational excitation on the diffuseness \cite{Gasques}. In contrast, our calculations for the $^{12}$C + $^{105,106}$Pd and $^{13}$C + $^{105,106}$Pd systems show a significant influence even though the target nuclei are spherical. 

In order to understand this difference in the role of channel coupling, we proceeded to find the parameters responsible for the influence of a vibrational excitation on the surface diffuseness parameter.  
We have first checked whether the deformation of the nucleus is responsible for this behavior. The table \ref{tab:beta} shows the deformation parameter $\beta$ for all the spherical target nuclei, $^{208}$Pb, $^{92}$Mo, and $^{105,106}$Pd. For comparison, the table also lists the deformation parameter for the deformed 
target nuclei, $^{186}$W and $^{170}$Er considered in Ref. \cite{Gasques}, which are estimated from the measured $B(E2)$ value \cite{Raman} with the radius parameter of $r_0=1.2$ fm. As is shown in the table, $^{208}$Pb and $^{92}$Mo have the deformation parameter which is significantly smaller than that for deformed 
nuclei. For those nuclei, the vibrational effect on the surface diffuseness parameter has been found to be marginal. In contrast, the Pd isotopes have a comparably large value of $\beta$ to the deformed nuclei, leading to a large channel coupling effect on the surface diffuseness parameter. 

\begin{table} 
\caption{\label{tab:beta}
The deformation parameter, $\beta$, the excitation energy, $E^*$, and the spin-parity 
for the vibrational state in 
spherical target nuclei involved in the coupled-channel calculations. 
The values for the odd-mass nucleus, $^{105}$Pd, are estimated by averaging those quantities for the neighbouring 
$^{104}$Pd and $^{106}$Pd nuclei. 
The values are taken from the references listed along with. 
The table also lists those quantities for the deformed nuclei studied in Ref. \cite{Gasques}, that is, 
$^{186}$W and $^{170}$Er. 
}
\begin{ruledtabular}
\begin{tabular}{ccccc}
Nucleus &   	$\beta $  & $E^*$ (MeV) & state & Ref. \\
\hline
\vspace{+0.075cm}
$^{208}$Pb & 0.111 & 2.615&  $3^-$ & \cite{kibedi} \\
\vspace{+0.075cm}
$^{92}$Mo  & 0.140 & 2.849&  $3^-$ & \cite{table} \\
\vspace{+0.075cm}
$^{105}$Pd & 0.219 & 0.534&  ``$2^+$'' & \cite{QEdata} \\
\vspace{+0.075cm}
$^{106}$Pd & 0.229 & 0.512& $2^+$ & \cite{QEdata} \\
\hline
$^{186}$W & 0.226 & 0.112& $2^+$ & \cite{QEdata} \\
$^{170}$Er & 0.336 & 0.0786& $2^+$ & \cite{QEdata} \\
\end{tabular}
\end{ruledtabular}
\end{table}





Naturally, one can expect that the energy of the excited state correlates with the deformation parameter. That is, the larger the deformation parameter is, the smaller the excitation energy will be. The table \ref{tab:beta} shows the energy of the first excited state of various nuclei considered in this paper as well as those in literature. It can be observed from the table that those nuclei which exhibit a large channel coupling effect on surface diffuseness parameter have a small 
excitation energy of the first excited state, i.e., $E^*<1$ MeV. In contrast, 
the first vibrational state of $^{208}$Pb is at 2.615 MeV, and thus the channel coupling effect is much smaller. 

Evidently, it is both $\beta$ and $E^*$ which are responsible for the influence of channel coupling effect on the surface diffuseness parameter. The nature of coupling scheme, that is, the rotational versus vibrational, is unimportant with respect to the influence on the diffuseness. 



\subsection{Role of Transfer Coupling}

In the coupled-channels calculation shown in the previous subsection, 
the couplings to the quadrupole vibrational states are considered. If these were the only dominant 
channels, one would expect that the extracted surface diffuseness parameters were similar among the 
systems.  

\begin{figure}[pb]
\includegraphics[width=.84\linewidth]{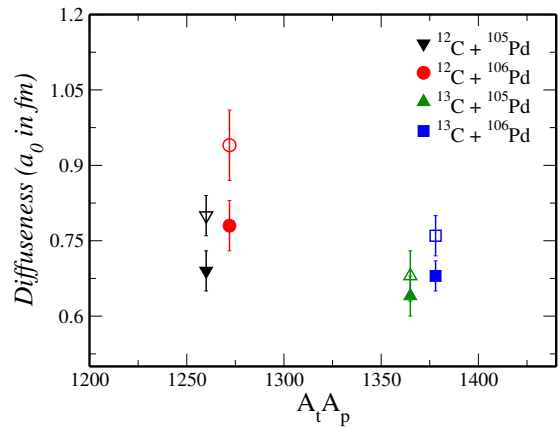}
\caption{\label{fig:apat}(Color online) 
The extracted surface diffuseness parameter for the $^{12,13}$C+$^{105,106}$Pd systems as a function of the 
product of masses of the projectile and target nuclei, $A_tA_p$. 
The open and filled symbols represent the results of the single-channel 
and the coupled-channel calculations, respectively.}
\end{figure}

Fig. \ref{fig:apat} 
shows the diffuseness parameter extracted with the single-channel (the open symbols) and the coupled-channels (the 
filled symbols) calculations as a function of the mass product of the projectile and the target 
nuclei, $A_tA_p$. 
It can be observed from the figure that the diffuseness parameter is reduced as the projectile isotope 
is changed from $^{12}$C to $^{13}$C (with the same target isotope) or the target isotope is changed 
from $^{106}$Pd to $^{105}$Pd (with the same projectile isotope). 
This gives us a hint that there could be an effect of channel coupling involved other than the 
collective quadrupole excitations. 
As we have discussed in Sec. II, the transfer channel is a promising candidate for this, since 
a large probability of transfer is expected for $^{13}$C due to the presence of the 
valence neutron. 


\begin{figure}[pt]
\includegraphics[width=.84\linewidth]{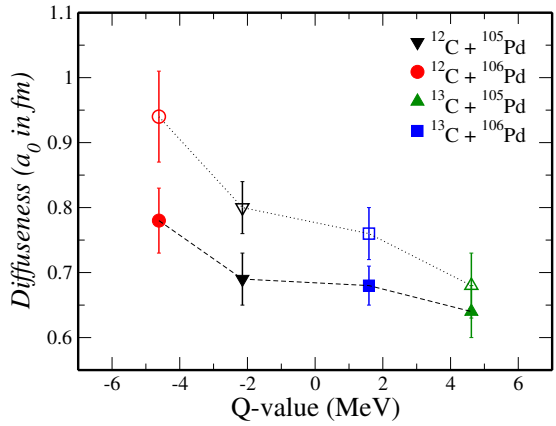}
\caption{\label{fig:a-trans}(Color online) 
Same as Fig. \ref{fig:apat}, but as a function of the $Q$-value for the neutron transfer channels. 
The lines are to guide the eye.}
\end{figure}

In order to investigate the role of transfer couplings, 
we have plotted in Fig. \ref{fig:a-trans} the optimum value of surface diffuseness parameter as a 
function of the Q-value for neutron transfer. 
Since for the present systems, the two-neutron (2n) transfer is a second step process, the most important transfer channel 
is a one-neutron (1n) transfer which may not be a general case. 
Hence, we have plotted the surface diffuseness as a function of the Q-value for the 1n transfer channels. 
For the 1n transfer channels, we have considered the $+1n$ and the $-1n$ channels for the $^{12}$C and 
$^{13}$C projectile nuclei, respectively. These would be most preferable transfer channels from the point of 
view of the transfer $Q$-value, although the $+1n$ channel may be equally important for the $^{13}$C+$^{105}$Pd 
system. 
It can be observed from the figure that, as a general trend, 
the surface diffuseness decreases  
as the transfer $Q$-value increases. This
might indicate that the difference in the surface diffuseness parameter between the $^{12}$C projectile 
and the $^{13}$C projectile 
could be attributed to the influence of neutron transfer coupling. 
It would be an intriguing future work to confirm this conjecture by carrying out coupled-channels 
calculations including both the collective excitations and the neutron transfer channels, although it is beyond 
the scope of the present paper.

\section{Summary}

The value of surface diffuseness parameter in the internuclear potential for the 
$^{12}$C + $^{105,106}$Pd and $^{13}$C + $^{105,106}$Pd systems has been extracted from the measured 
quasi-elastic scattering cross sections at a backward angle. 
To this end, both the single-channel calculations and the coupled-channels calculations including the 
quadrupole vibrational excitations in the target nuclei have been carried out. 
Even though the systems studied involve spherical nuclei, the comparison of the coupled-channels calculations 
with the single-channel calculations revealed the reduction in the values of the diffuseness parameter. 
A similar reduction had been 
reported earlier for deformed systems due to the rotational coupling. Evidently, the conclusion in Refs. \cite{Gasques,Evers08,Jiang14} that the vibrational
excitation has a marginal effect on the surface diffuseness parameter
is not a general one, but instead, the effect becomes significant
even for the vibrational coupling when the coupling is strong enough. We have argued that the reduction in the extracted diffuseness parameter 
is due to the strong coupling to the low-lying collective states, and the nature of coupling is not 
important. That is, the reduction can be observed both for the rotational and the vibrational couplings as 
long as the coupling strength is large enough. 
Furthermore, a discussion has been made in order to understand the influence of the 
transfer coupling on the surface diffuseness parameter. 
We have observed that the surface diffuseness parameter gets smaller as the transfer $Q$-value increases. 
This implies that the surface diffuseness 
tends to be smaller when the transfer coupling is stronger. 
It would be an interesting future work to confirm whether this trend holds in other systems as well. 
For that purpose, it would be 
interesting also to perform coupled-channels calculations including the transfer degree of freedom 
and carry out systematic studies in order to clarify the interplay between the collective excitations and 
the transfer couplings.

\section*{Acknowledgment}
One of the authors, G. K. acknowledges the University Grants Commission (UGC), New Delhi for providing the financial assistance for this work.

%


\begin{thebibliography}{50}
\bibitem{Steadman} S.G. Steadman and M.J. Rhoades-Brown, Annu. Rev. Nucl. Part. Sci. {\bf 36}, 649 (1986).

\bibitem{Dasgupta1} M. Dasgupta, D.J. Hinde, N. Rowley, and A.M. Stefanini, Annu. Rev. Nucl. Part. Sci. {\bf 48}, 401 (1998).

\bibitem{BT98} A.B. Balantekin and N. Takigawa, Rev. Mod. Phys. {\bf 70}, 77 (1998).

\bibitem{HT12} K. Hagino and N. Takigawa, Prog. Theor. Phys. {\bf 128}, 1001 (2012).

\bibitem{Back14} B.B. Back, H. Esbensen, C.L. Jiang, and K.E. Rehm, Rev. Mod. Phys. {\bf 86}, 317 (2014).

\bibitem{Stefanini} A.M. Stefanini, D. Ackermann, L. Corradi, J.H. He, G. Montagnoli, S. Beghini, F. Scarlassara, and G.F. Segato, Phys. Rev. C {\bf 52}, R1727 (1995).

\bibitem{Dasgupta2} M. Dasgupta, D.J. Hinde, J.R. Leigh, and K. Hagino, Nucl. Phys. {\bf A630}, 78 (1998).

\bibitem{Hagino1} K. Hagino, M. Dasgupta, I.I. Gontchar, D.J. Hinde, C.R. Morton, and J.O. Newton, in Proceedings of the Fourth Italy-Japan Symposium on Heavy-Ion Physics, Tokyo, edited by K. Yoshida et al. (World Scientific, Singapore, 2002), pp. 87; nucl-th/0110065.

\bibitem{Gonttchar} I.I. Gonttchar, D.J. Hinde, M. Dasgupta, and J.O. Newton, Phys. Rev. C {\bf 69}, 024610 (2004).

\bibitem{wood} R.A. Broglia and A. Winther, Heavy Ion Reactions, in Frontiers in Physics Lecture Note Series (Addison-Wesley, Redwood City, CA, 1991), Vol. 84.

\bibitem{elastic} M. Lozano and G. Madurga, Nucl. Phys. {\bf A334}, 349 (1980).

\bibitem{newton1} J.O. Newton, R.D. Butt, M. Dasgupta, D.J. Hinde, I.I. Gontchar, C.R. Morton, and K. Hagino, Phys. Lett. {\bf B586}, 219 (2004).

\bibitem{newton2} J.O. Newton, R.D. Butt, M. Dasgupta, D.J. Hinde, I.I. Gontchar, C.R. Morton, and K. Hagino, Phys. Rev. C {\bf 70}, 024605 (2004).

\bibitem{Hagino2} K. Hagino, T. Takehi, A.B. Balantekin, and N. Takigawa, Phys. Rev. C {\bf 71}, 044612 (2005).

\bibitem{Washiyama} K. Washiyama, K. Hagino, and M. Dasgupta, Phys. Rev. C {\bf 73}, 034607 (2006).

\bibitem{92Mo} D.S. Monteiro, J.M.B. Shorto, J.F.P. Huiza, P.R.S. Gomes, and E. Crema, Phys. Rev. C {\bf 76}, 027601 (2007).

\bibitem{Gasques} L.R. Gasques, M. Evers, D.J. Hinde, M. Dasgupta, P.R.S. Gomes, R.M. Anjos, M.L. Brown, M.D. Rodrguez, R.G. Thomas, and K. Hagino, Phys. Rev. C {\bf 76}, 024612 (2007).
 
\bibitem{Evers08} M. Evers, M. Dasgupta, D. Hinde, L. Gasques, M. Brown, R. Rafiei, and R. Thomas, Phys. Rev. C {\bf 78}, 034614 (2008). 

\bibitem{Jiang14} C.L. Jiang, K.E. Rehm, B.B. Back, H. Esbensen, R.V.F. Janssens, A.M. Stefanini, and G. Montagnoli, Phys. Rev. C {\bf 89}, 051603(R) (2014). 

\bibitem{QEdata} O.A. Capurro, J.E. Testoni, D. Abriola, D.E. DiGregorio, G.V. Martı, A.J. Pacheco, M.R. Spinella, and E. Achterberg, Phys. Rev. C {\bf 62}, 014613 (2000).

\bibitem{Qvalue} R.A. Broglia, C.H. Dasso, S. Landowne, and A. Winther, Phys. Rev. C {\bf 27}, R2433 (1983).

\bibitem{ccfull} K. Hagino, N. Rowley, and A.T. Kruppa, Comput. Phys. Commun. {\bf 123}, 143 (1999).

\bibitem{bass} R. Bass, Phys. Rev. Lett. {\bf 39}, 265 (1977).

\bibitem{Hagino3} K. Hagino, N. Takigawa, and S. Kuyucak, Phys. Rev. Lett. {\bf 79}, 2943 (1997).

\bibitem{eff1} H. Timmers, J.R. Leigh, M. Dasgupta, D.J. Hinde, R.C. Lemmon, J.C. Mein, C.R. Morton, J.O. Newton and N. Rowley, Nucl. Phys. {\bf A584}, 190 (1995).

\bibitem{eff2} K. Hagino and N. Rowley, Phys. Rev. C {\bf 69}, 054610 (2004).

\bibitem{table} T. Nakagawa et al., "Japanese evaluated nuclear data library, version 3 revision-2; JENDL-3.2", J. Nucl. Sci. Technol. {\bf 32}, 1259 (1995).

\bibitem{kibedi} T. Kib\'edi and R.H. Spear, Atomic Data and Nuclear Data Tables {\bf 80}, 35–82 (2002). 

\bibitem{Raman} S. Raman, C.W. Nestor, JR., and P. Tikkanen, At. Data Nucl. Data Tables {\bf 78}, 1 (2001). 


\end{thebibliography}
\end{document}